\documentclass[%
 reprint,
 amsmath,amssymb,
 aps,
]{revtex4-1}

\usepackage[dvips]{graphicx}% Include figure files
\usepackage{dcolumn}% Align table columns on decimal point
\usepackage{bm}% bold math

\usepackage{color,graphicx}
\usepackage[colorlinks=true,urlcolor=blue,citecolor=blue,linkcolor=blue]{hyperref}
\usepackage{epstopdf}

\newcommand{\WV}[1]{\textcolor{black}{#1}}
\newcommand{\BK}[1]{\textcolor{black}{#1}}
\newcommand{\BKTWO}[1]{\textcolor{black}{#1}}
\newcommand{\BKTHREE}[1]{\textcolor{black}{#1}}

\begin{document}
\preprint{APS/123-QED}

\title{%Optimal 
Visualizing nonclassical effects in phase space}% Force line breaks with \\

\author{B. K\"uhn}
 
 \email{benjamin.kuehn2@uni-rostock.de}

\author{W. Vogel}

\affiliation{
 Arbeitsgruppe Quantenoptik, Institut f\"ur Physik, Universit\"at Rostock, D-18051 Rostock, Germany
}

\date{\today}

\begin{abstract}
Nonclassicality filters provide a universal method to visualize the nonclassicality of arbitrary quantum states of light through negativities of 
a regularized Glauber-Sudarshan $P$~function, also denoted as nonclassicality quasiprobability.
%The method is based on nonclassicality filters as introduced in 
%[T. Kiesel and W. Vogel, Phys. Rev. A \textbf{82}, 032107 (2010)]. 
Such filters are introduced and analyzed for optimizing the experimental certification of nonclassical effects. An analytic filter is constructed which preserves the full information on the quantum state. For balanced homodyne detection, the number of data points is analyzed to get the negativities of the nonclassicality quasiprobability with high statistical significance. The method is applied to different scenarios, such as phase randomized squeezed vacuum states, single-photon-added thermal states, and heralded state engineering with array detectors. The generalization to visualize quantum correlations of multimode radiation fields is also considered.\\ 
\begin{description}
\item[PACS numbers]03.65.Wj, 42.50.Dv, 03.65.Ta
\end{description}
\end{abstract}

\pacs{Valid PACS appear here}
\maketitle

\section{Introduction}
The visualization of features of a physical system which have no counterpart in classical statistical physics has attracted increasing interest. The most prominent nonclassical phenomenon in multidimensional systems is entanglement, which is a key resource for quantum technologies such as quantum computation %Lia1} 
and quantum teleportation~\cite{Nielsen,HorodeckiRev,Ben1}. However, \WV{even a single-mode harmonic oscillator exhibits a variety of nonclassical properties. %Especially for optical systems, 
It is well known} that the generation of an entangled state of the radiation field in the output beams of a beam splitter necessarily requires one input field to be prepared in a nonclassical state~\cite{ahar,kim,xiangbin,WEP03,JLC13}. Hence it is a subject of great importance to have powerful and general criteria to certify the nonclassicality of a given quantum state. 

An established definition of nonclassicality of the radiation field is based on the Glauber-Sudarshan representation of the quantum state~\cite{Gla1,Sud1},
\begin{equation}
\hat\rho=\int d^2\alpha\,P(\alpha)|\alpha\rangle\langle\alpha|,
\label{eq:P-repr}
\end{equation}
in terms of coherent states $|\alpha\rangle$, which are closely related to the classical behavior.
If the $P$~function has the properties of a classical probability density, then Eq.~(\ref{eq:P-repr}) corresponds to a classical mixture of the (almost classical)
coherent states. Such quantum states are called classical ones, for nonclassical states $P(\alpha)$ fails to have the properties of a probability density~\cite{Tit1,Man1}.
For a large number of quantum states the Glauber-Sudarshan $P$~function can have negativities. Such states exhibit quantum effects arising from quantum superpositions of coherent states~\cite{Geh1}; their properties are unknown in classical physics.
%Therefore, they are called nonclassical ones.

In general, $P(\alpha)$ is highly singular, so that it is often impossible to reconstruct it from experiments in order to certify the nonclassicality of a given quantum state. Even if the $P$~function is well behaved~\cite{Ag-Tara}, its experimental reconstruction requires some \textit{a priori} knowledge of the set of possible states in the considered physical system. In such cases, a cutoff of the Fourier transform of the characteristic function must be properly chosen~\cite{Kie2} in order to suppress the experimental sampling noise effects.

A universal method for verifying nonclassicality introduces a regularized $P$~function by applying a so-called nonclassicality filter~\cite{Kie1}. Negativities of this regularized function are proof of the nonclassicality of the state. In contrast to the Glauber-Sudarshan $P$~function, it is a regular function for any quantum state and can, in general, be reconstructed from experimental data. This filtering procedure is a very efficient technique since it only requires the optimization of three real parameters. Other nonclassicality criteria, for example nonclassicality conditions~\cite{Ric2} derived from the Bochner criterion~\cite{Boc1} or criteria using normally ordered moments~\cite{Ag-Tara,Ag,Shc1}, in general require an infinite number of conditions to certify nonclassicality. Special signatures of nonclassicality, such as a sub-Poisson photon statistics~\cite{Dal1}, quadrature squeezing~\cite{Slusher,Lin1}, or negative values of the Wigner function~\cite{Wineland}, identify only the 
nonclassicality of a subset of all nonclassical states.
The practicability of the nonclassicality filtering has been demonstrated in different experiments~\cite{Kie5,Kie-Schn,Kie-Polz}.
Little attention has been paid so far to the following question: which filter needs a minimal number of experimental data points to obtain significant negativities of the experimentally reconstructed regularized $P$~function?

% By contrast other nonclassicality criteria, as for example the Bochner criterion~\cite{Boc1,Ric2}, or criteria which use normally ordered moments~\cite{Shc1}, in the worst case require the evaluation of an infinite number of conditions to prove nonclassicality. Nonclassical signatures such as a sub-Poisson photon statistics~\cite{Dal1}, quadrature squeezing~\cite{Lin1} or negative values of the Wigner function~\cite{Cah1}, are only able to prove the nonclassicality of a subset of all nonclassical states.

In the present paper we study known \WV{nonclassicality filters and introduce additional ones.} Two types of such filters can even be given in an analytic form, which is advantageous for the practical application in experiments. We compare different filters to identify the optimal strategy for uncovering quantum effects of a given system. We analyze the \WV{number of experimental data points,} needed to achieve a high significance of the quantum effects of interest. The method is applied to 
lossy Fock states, $n$-photon-added thermal states, and fully dephased squeezed vacuum states. The \WV{possibility to uncover, with our approach, quantum correlations of multimode radiation fields is also studied and applied to an exampl\BKTWO{e} two-mode quantum correlation.}

%paper, we will pursue this question by investigating the quality of various nonclassicality filters for the application to different states.

This article is structured as follows. In Sec.~\ref{ch:nonclassicalityfilters}, we \WV{compare the properties of different nonclassicality filters, and we construct 
%two other 
\BKTWO{a filter which is} analytical and invertible. Different nonclassicality filters are applied to several standard quantum states in Sec.~\ref{ch:analysisstandard} to answer the question of which filter 
needs the smallest number of data points to verify nonclassicality via balanced homodyne detection.
%\BK{The results are discussed in Section~\ref{ch:discussion}.} %In Section~\ref{ch:heraldedstateengineering}, the same analysis is made for output states of a heralded state preparation setup. 
A multimode nonclassicality quasiprobability is considered in Sec.~\ref{ch:twomode} and is applied to visualize two-mode quantum correlations. A summary  
%our results 
and some conclusions are given} in Sec.~\ref{ch:conclusion}. 

\section{Nonclassicality filters}
\label{ch:nonclassicalityfilters}
\subsection{Definition and properties}
The Glauber-Sudarshan \WV{$P$~function, 
\begin{equation}
P(\alpha)=\dfrac1{\pi^2}\int d^2\beta\, e^{\alpha\beta^*-\alpha^*\beta}\,\Phi(\beta),
\end{equation}
can be determined as the Fourier transform
of the characteristic function $\Phi(\beta)$. In general, the latter does not tend to zero for $|\beta|\to\infty$, which often} leads to a singular $P$~function. 
The regularization of this function is based on the multiplication of the characteristic function with a filter $\Omega_w(\beta)$ before the Fourier transform is carried out. This leads to the regularized $P$~function~\cite{Kie1},
\begin{equation}\label{eq:filterP}
P_\Omega(\alpha;w)=\dfrac1{\pi^2}\int d^2\beta\, e^{\alpha\beta^*-\alpha^*\beta}\,\Omega_w(\beta)\,\Phi(\beta),
\end{equation}
which is also referred to as filtered $P$~function or nonclassicality quasiprobability. \WV{The filter function is controlled by the positive parameter $w$, the filter width. As the value of $w$ increases, the structures of the filtered $P$~function typically become sharper.} For classical states any choice of $w$ yields a nonnegative function $P_\Omega(\alpha;w)$. \WV{On the other hand, for  each nonclassical state a filter width} exists such that $P_\Omega(\alpha;w)$ has negativities, directly revealing the nonclassicality. 

The filter function $\Omega_w(\beta)$, appearing in Eq.~\eqref{eq:filterP}, has to be chosen in a special way. It is important that the filtered quasiprobability $P_\Omega$ can visualize the quantumness of any nonclassical state. In addition, the filter must suppress the sampling noise of the measured data.

\textit{Condition 1}. $\Omega_w(\beta)e^{|\beta|^2/2}$ is square integrable for all $w>0$. This guarantees that the nonclassicality quasiprobability is a regular function for any quantum state. Moreover, it fully suppresses the experimental sampling noise in the characteristic function of the $P$~function.

\textit{Condition 2}. Negative values of $P_\Omega(\alpha;w)$ should only arise from the nonclassicality of the state and not from the filter procedure itself. Therefore, the Fourier transform of $\Omega_w(\beta)$ has to be nonnegative for all $w>0$.

\textit{Condition 3}. $\Omega_w(\beta)=\Omega_w^*(-\beta)$ and $\Omega_w(0)=1$, so $P_\Omega$ is a real function and $\int d^2\alpha\,P_\Omega(\alpha;w)= 1$.

\textit{Condition 4}. If $w$ tends to infinity, $P_\Omega$ has to approach the original Glauber-Sudarshan $P$~function, which implies that $\lim_{w\to\infty}\Omega_w(\beta)=1$ must hold.

A function satisfying \textit{Conditions 1} - \textit{4} is referred to as a nonclassicality filter~\cite{Kie1}.
If the quantum state is completely known, i.e., if it is, for example, given by its density operator or its characteristic function, it is useful to add another condition.

\textit{Condition 5}. $\Omega_w(\beta)\neq 0$ for all $\beta\in\mathbb{C}$ and $w>0$. This ensures that the regularized $P$~function represents all quantum states uniquely, without any loss of information. If a nonclassicality filter satisfies this condition, we refer to it as an invertible nonclassicality filter. For sampling the regularized $P$~function from \WV{experiments, however, this condition is dispensable since quantum information is already lost due to the finite number of 
recorded} data points.

It is advantageous to use radial symmetric filters since the sampling formulas for the reconstruction of the regularized $P$~function from quadrature or photon number data are easier to handle in this case. Therefore, we add a further requirement.

\textit{Condition 6}. $\Omega_w(\beta)=\Omega_w(|\beta|)$. Together with \textit{Condition 3}, one infers that $\Omega_w(\beta)$ is a real function in this case.

\subsection{Presently known nonclassicality filters}
In the following we will give a brief overview of presently known nonclassicality filters, and we will introduce another analytical nonclassicality filter.
One possibility to construct nonclassicality filters is based on the autocorrelation function,
\begin{equation}\label{eq:autq}
\Omega^{(q)}_w(\beta)=\dfrac{q\,2^{2/q-1}}{\pi\Gamma\left(2/q\right)}\,\int d^2\gamma\,e^{-|\gamma|^q}\,e^{-|\beta/w+\gamma|^q}, 
\end{equation}
where $\Gamma(x)$ is the gamma function. According to Ref.~\cite{Kie1}, function~\eqref{eq:autq} is a nonclassicality filter for $q>2$. The corresponding quasiprobability then reads
\begin{equation}
P^{(q)}_\Omega(\alpha;w)=\dfrac1{\pi^2}\int d^2\beta\,e^{\alpha\beta^*-\alpha^*\beta}\,\Omega^{(q)}_w(\beta)\,\Phi(\beta). 
\end{equation}
In the case $q=2$ one recovers the well-known $s$-parametrized quasiprobabilities, such as the Wigner function ($w=1$) or the Husimi $Q$~function ($w=1/\sqrt{2}$)~\cite{Cah1}. For $w>1$, however, the quasiprobability can still be highly singular for some states, which renders an experimental reconstruction impossible. Thus, this Gaussian filter violates \textit{Condition 1} and is, therefore, not a nonclassicality filter.

In the limiting case $q=\infty$, we obtain the analytical expression
\begin{equation}\label{eq:FilterInfty}
\Omega_w^{(\infty)}(\beta)=\dfrac{2}{\pi}\left[\arccos\left(\dfrac{|\beta|}{2w}\right)-\dfrac{|\beta|}{2w}\sqrt{1-\dfrac{|\beta|^2}{4w^2}}\right]\mathrm{rect}\left(\dfrac{|\beta|}{4w}\right),
\end{equation}
with
\begin{equation}
\mathrm{rect}(x)=
\begin{cases}
  1 & \text{if }|x|\leq 1/2,\\
  0 & \text{otherwise}.
\end{cases}
\end{equation}
Under more general conditions, for $2<q<\infty$, the filters $\Omega_w^{(q)}(\beta)$ must be determined numerically. \WV{These filters are invertible, and hence, they preserve the full information on the quantum state. Contrary to this, $\Omega_w^{(\infty)}
(\beta)$ in Eq.~\eqref{eq:FilterInfty} becomes zero for $|\beta|/w>2$.} This implies some partial loss of information on the detailed structure of the quantum states under study~\cite{Kie4}.
\begin{figure}[h]
\hbox{\hspace{0.5cm}\includegraphics[clip,scale=0.40]{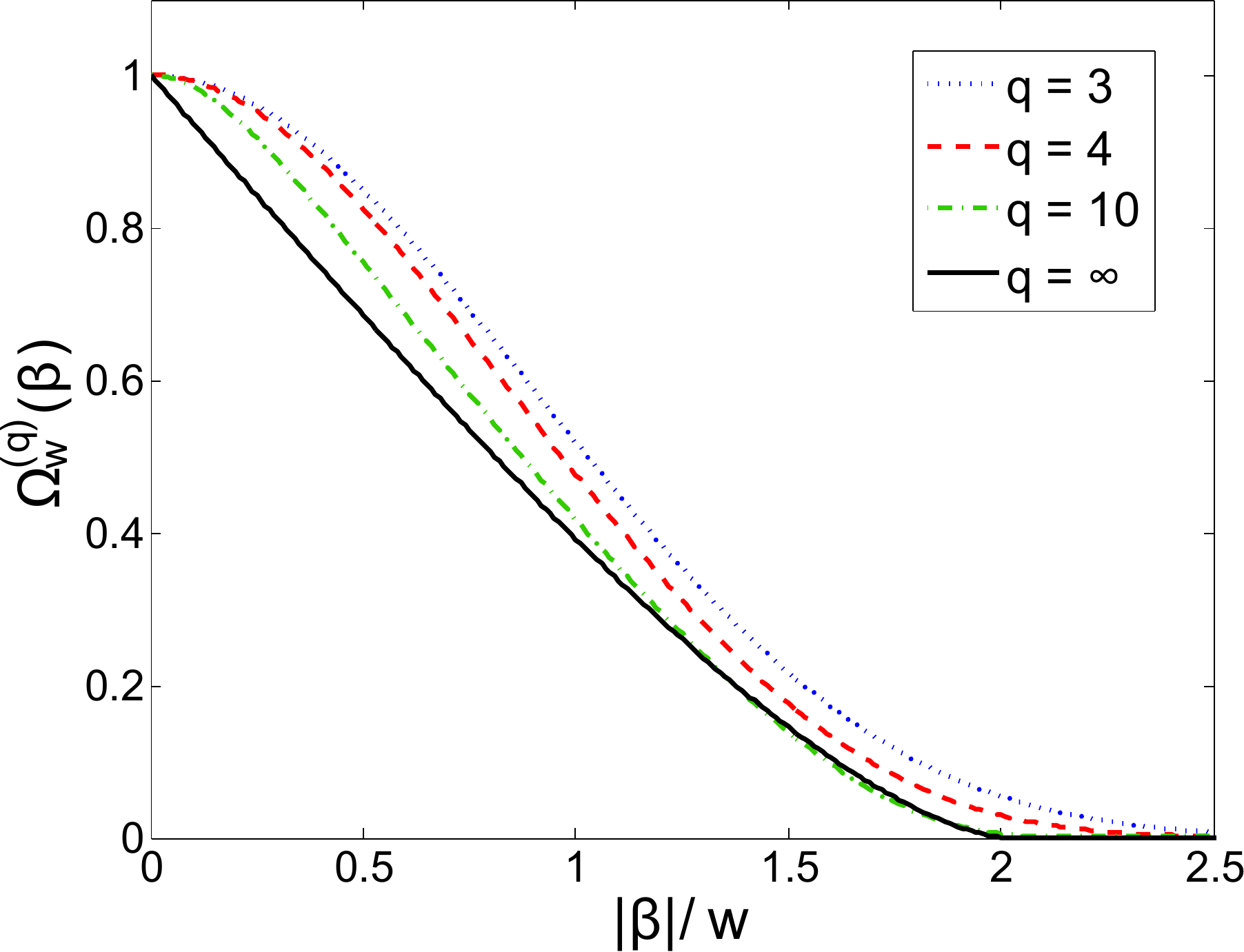}}
\vspace{0.3cm}
\hbox{\hspace{0.32cm}\includegraphics[clip,scale=0.398]{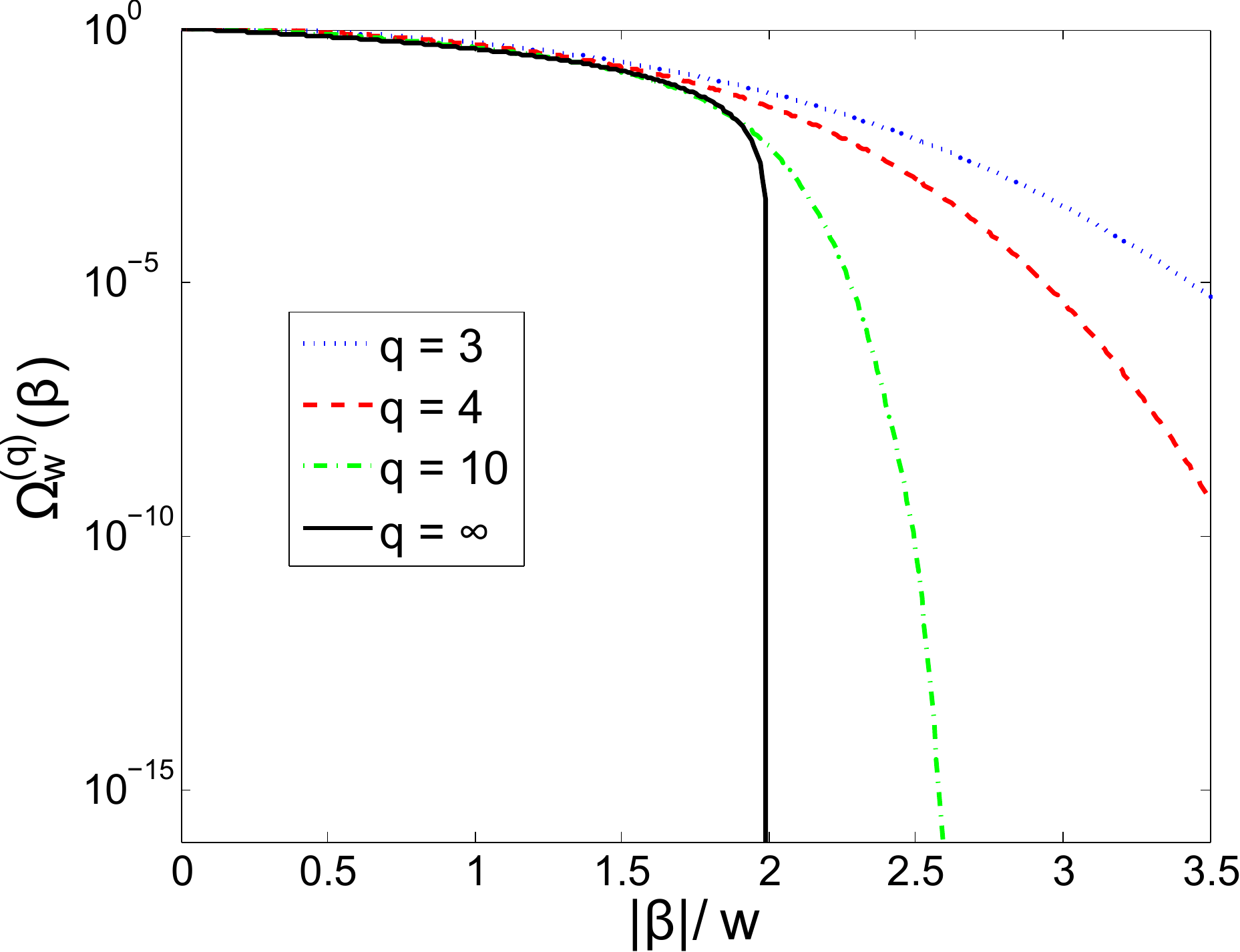}}
\caption{(Color online) Filter $\Omega_w^{(q)}(|\beta|)$ for different values of the parameter $q$. (top) Vertical axis linear. (bottom) Vertical axis logarithmic.}
\label{fig:Autq}
\end{figure}
Figure~\ref{fig:Autq} shows the linear and logarithmic shapes of these filters along the radial direction in phase space, for various values of $q$. The parameter $q$ mainly controls the decay of $\Omega_w^{(q)}(\beta)$ for $|\beta|/w>2$, which becomes stronger with increasing $q$. In the following sections we will see that this $q$-dependent decay behavior exceedingly determines the statistical significance of the negativities of the experimentally reconstructed regularized $P$~functions. An important question arises: for which $q$ values does a minimal number of experimental data points suffice to reach a reasonable statistical significance of the visualized nonclassicality? We will deal with this question in Sec.~\ref{ch:analysisstandard}. %and~\ref{ch:heraldedstateengineering}.

\subsection{Analytical invertible nonclassicality filter}
\label{sec:anafilter}
The previous examples motivate us \BKTHREE{to} search for a nonclassicality filter that is both analytical and invertible, whose construction has not been possible so far. \WV{Here we introduce}
%We succeeded to find 
a simple analytic example of such a filter which fulfills \textit{Conditions 1}--\textit{6}. It is given by
\begin{equation}\label{eq:anafilter}
\Omega_w(\beta;s,C)=\exp\left[-\left(\dfrac{|\beta|}{w}+C\right)^s+C^s\right],
\end{equation}
controlled by the three real parameters, $w$, $s$, and $C$. In the Appendix we prove that
\begin{equation}\label{eq:sgreater2}
s>2
\end{equation}
ensures \textit{Condition 1}. The proof of the nonnegativity of the Fourier transform of~\eqref{eq:anafilter} is based on a theorem by Askey about positive-definite functions~\cite{Ask1}. Consider a radial continuous function $f(\boldsymbol{x})=\varphi(t)$, with $\boldsymbol{x}\in\mathbb{R}^n$ and $t=\|\boldsymbol{x}\|$ being the Euclidean norm of $\boldsymbol{x}$. It is a characteristic function in $\mathbb{R}^n$; that is the Fourier transform of a probability measure if the following requirements are fulfilled: $\varphi(0)=1$, $\lim_{t\to\infty}\varphi(t)=0$, and $(-1)^k\varphi^{(k)}(t)$ is convex for $k=\lfloor n/2 \rfloor$, where $k$ is the greatest integer less than or equal to $n/2$. Here $\varphi^{(k)}$ denotes the $k$th derivative of $\varphi$ with respect to $t$. In the special case of a single mode the phase space is two-dimensional ($n=2$). Accordingly, the conditions of the theorem have to be satisfied for $k=1$. Therefore, the filter~\eqref{eq:anafilter} has to satisfy
\begin{equation}\label{eq:thirdderivative}
-\dfrac{d^3}{d|\beta|^3}\Omega_w(\beta;s,C)\geq 0
\end{equation}
for all $|\beta|$. Inserting Eq.~\eqref{eq:anafilter} into inequality~\eqref{eq:thirdderivative} yields the condition
\begin{equation}\label{eq:cmincond}
C\geq C_{\mathrm{min}}(s)=\left(\dfrac{3(s-1)+\sqrt{1-6s+5s^2}}{2s}\right)^{1/s}
\end{equation}
(see the Appendix). The maximum is $\max_{s>2} C_{\mathrm{min}}(s)\approx 1.24541$. If the parameters $s$ and $C$ are chosen such that~\eqref{eq:sgreater2} and~\eqref{eq:cmincond} are fulfilled, then $\Omega_w(\beta;s,C)$ is a nonclassicality filter. Therefore, we choose $C=1.3$, which applies to all $s$ values with $s>2$.

\begin{figure}[h]
\includegraphics[clip,scale=0.43]{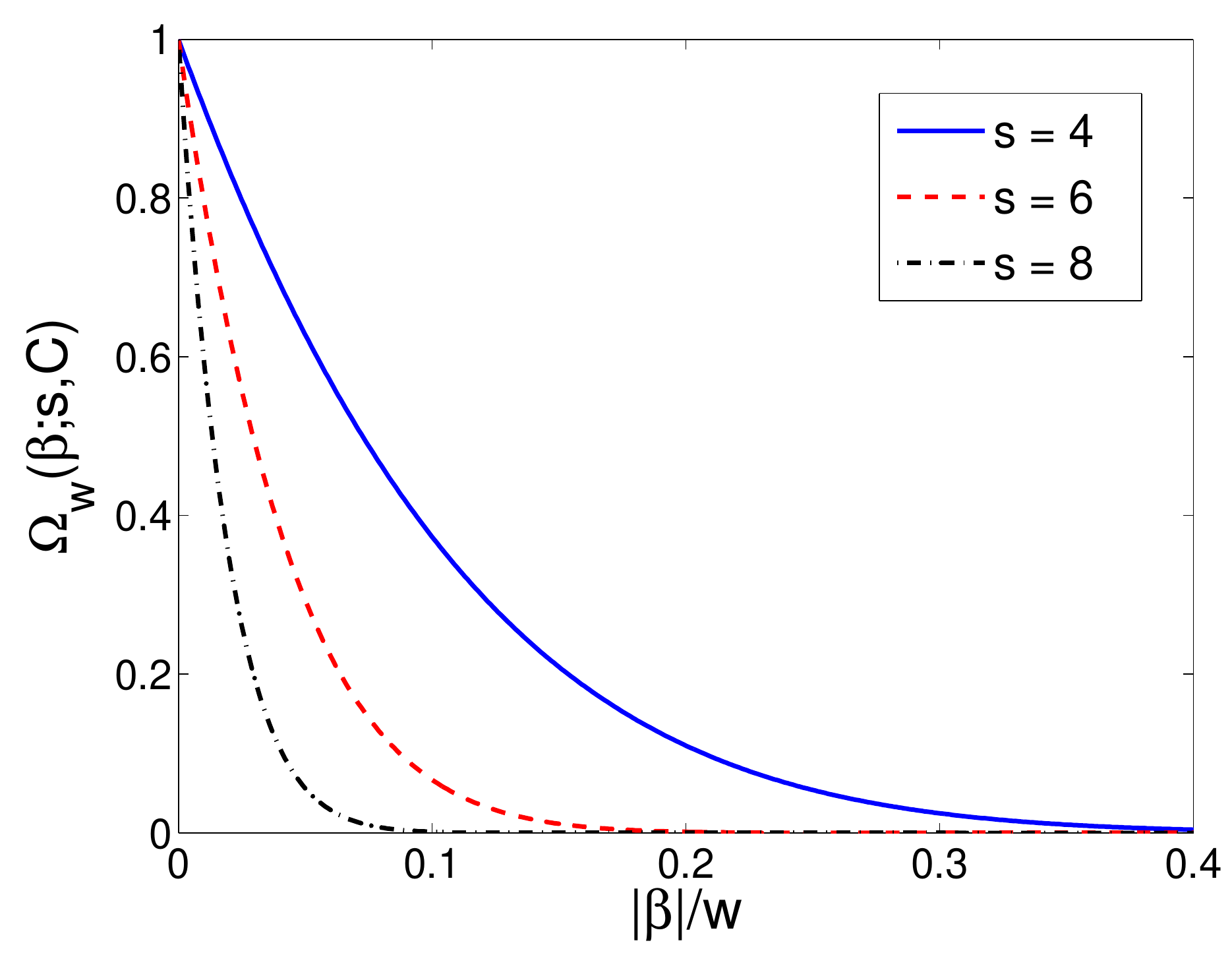}
\caption{(Color online) Filter $\Omega_w(|\beta|;s,C)$ for $C=1.3$ and different values of the parameter $s$.}
\label{fig:AnaFilter}
\end{figure}

Since the analytical filter in Eq.~\eqref{eq:anafilter} is invertible, i.e., $\Omega_w(\beta)\neq 0$ for all $\beta\in\mathbb{C}$, it preserves the full information about the quantum state, contrary to the analytical filter $\Omega_w^{(\infty)}(\beta)$ in Eq.~\eqref{eq:FilterInfty}. The radial shape of $\Omega_w(\beta;s,C)$ is given in Fig.~\ref{fig:AnaFilter} for various values of the parameter $s$. For direct sampling of the regularized $P$~function from experimental quadrature data, the filter $\Omega_w^{(\infty)}(\beta)$ is more useful than
$\Omega_w(\beta;s,C)$. In fact, orders of magnitude more data points are required for the latter compared with the former when a certain statistical significance of the negativity of $P_\Omega(\alpha;w)$ is required. 

\begin{figure}[h]
\includegraphics[clip,scale=0.37]{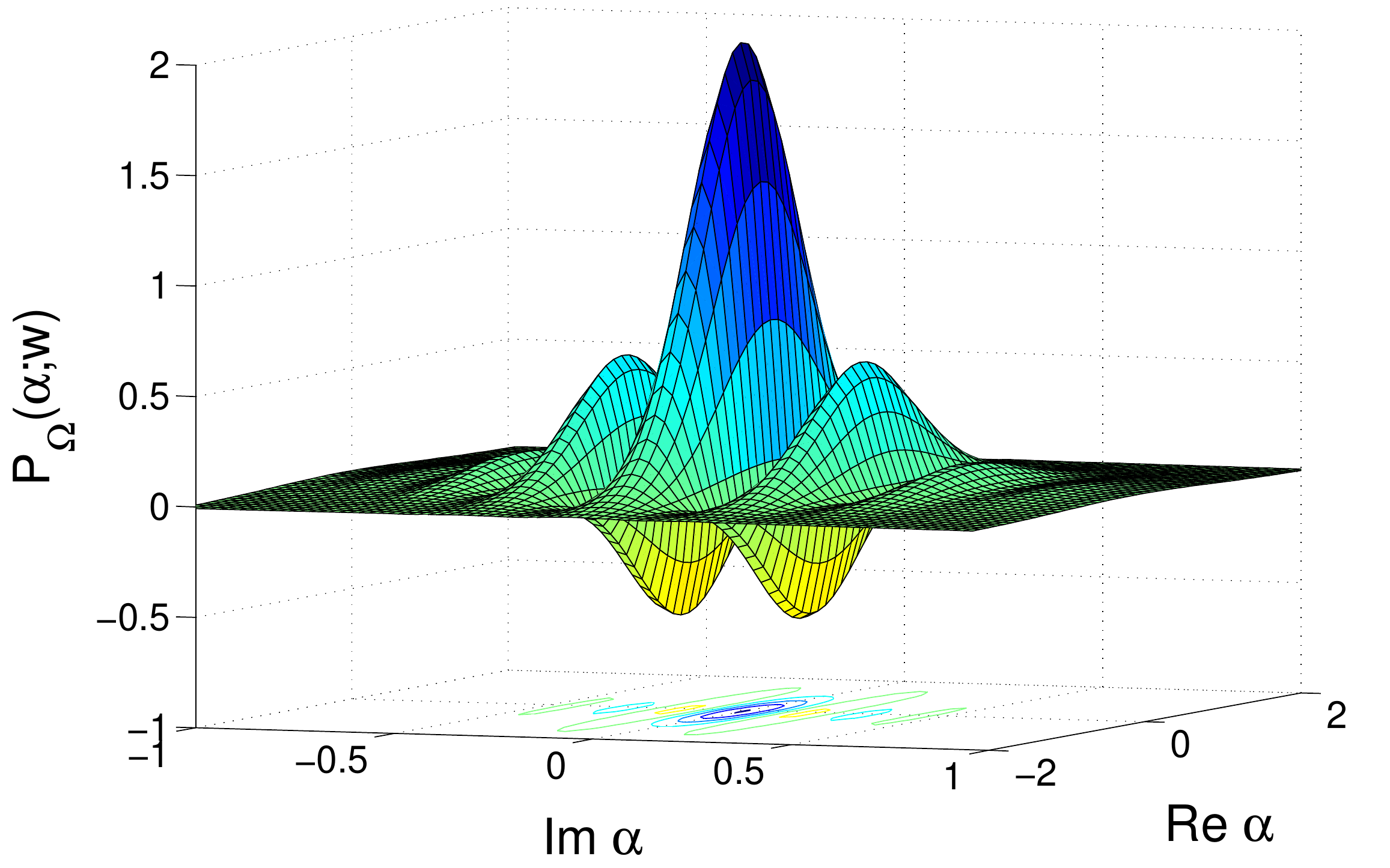}
\caption{(Color online) $P_\Omega(\alpha;w)$ of a squeezed vacuum state with orthogonal quadrature variances $V_p=2.0$ and $V_x=0.5$. The filter $\Omega_w(\beta;s,C)$ is applied, with $w=9.9$,  $C=1.3$, and $s=4$.}
\label{fig:Sqana} 
\end{figure}

For analyzing nonclassicality of quantum states which are given by theory in an analytical form, the nonclassicality filter $\Omega_w(\beta;s,C)$ is very useful since it is simple to implement. It requires neither numerical effort nor storage space. Even in such cases it is often unclear whether the state is nonclassical or not, as a strongly singular $P$~function may hide these effects. Moreover, the resulting regularized $P$~function is a complete representation of the quantum state under study. Hence, this filter is very useful for simulating and optimizing experiments which aim at preparing quantum states with certain types of nonclassical effects. \BKTHREE{Based on the analytical and invertible filter introduced in Eq.~\eqref{eq:anafilter}, we give in Fig.~\ref{fig:Sqana} the example of a nonclassicality quasiprobability of a squeezed vacuum state based on the theoretical characteristic function.} Distinct negativities appear, which clearly visualize the nonclassicality of this state. 

\subsection{Direct sampling of nonclassicality quasiprobabilities by balanced homodyne detection}
In the following we will study the possibility of direct sampling of nonclassicality quasiprobabilities from data points recorded by standard balanced homodyne detection (BHD)~\cite{Yue1}; for details see, e.g., the review in Ref.~\cite{We99} and references therein. \BKTHREE{In BHD the light field to be investigated is combined with a strong reference light field at a beam splitter, where the phase difference $\varphi$ of both beams is adjustable. The intensities of the output fields are recorded by two photodetectors. The correlated difference of the measured electric currents of these detectors yields the quadrature $x$ according to the quadrature distribution $p(x;\varphi)$ at phase $\varphi$. In the following we assume that quadrature-phase pairs $(x_j,\varphi_j)$ at different times $t_j$ are recorded. The phases $\varphi_j$ follow a uniform distribution in the interval $[0,\pi)$.}

Recently a method has been proposed which yields an estimate of the regularized $P$~function from a number $N$ of sampled %quadrature data points $x_j$ at the phases $\varphi_j$~\cite{Kie-Schn}. 
\BKTHREE{quadrature-phase pairs}~\cite{Kie-Schn}.
\BKTHREE{An estimate yields a value for an unknown quantity based on a given set of data points. This value approaches the actual value of this quantity if the number of data points tends to infinity.} It was shown that a proper estimate \WV{$\tilde{P}_\Omega$ of $P_\Omega(\alpha;w)$} is given by the average
\begin{equation}\label{eq:PEst}
\WV{\tilde{P}_\Omega(\alpha;w)=}\dfrac1{N}\sum_{j=1}^N f_\Omega(w,\Lambda_{j,\alpha})\BKTHREE{.}
\end{equation}
\BKTHREE{It is an unbiased estimate; that is, its expectation value is equal to the quantity $P_\Omega(\alpha;w)$ to be estimated.}
\BKTWO{The} pattern function reads
\begin{equation}
\label{eq:pattern}
f_\Omega(w,\Lambda_{j,\alpha})=\dfrac{2}{\pi}\int_0^\infty db\,b\,e^{b^2/2}\Omega_w(b)\,\cos\left(\Lambda_{j,\alpha}\,b\right).
\end{equation}
It depends on \WV{the filter width $w$, }\BKTWO{the radial symmetric nonclassicality filter $\Omega_w(\beta)=\Omega_w(b)$, with $b=|\beta|$, }\WV{and
\begin{equation}
\Lambda_{j,\alpha}=x_j+2|\alpha|\sin\left[\arg(\alpha)+\varphi_j-\pi/2\right],
\end{equation}
which contains} the phase-space point $\alpha$ and the measured quadrature $x_j$ for phase~$\varphi_j$. 
% \WV{The strong point of this method is that it yields the nonclassicality quasiprobability locally in phase-space for any point $\alpha$, which is more direct
% than state reconstruction methods based on inverse Radon transform or maximum likelihood methods~\cite{We99,Smi1,Hradil}.}
For more details, including the numerically efficient computation of the pattern function, we refer to the Supplemental Material of~\cite{Kie-Schn}.

An estimate for the corresponding variance of the sampled quasiprobability in Eq.~\eqref{eq:PEst} is
\BK{
\begin{eqnarray}
\label{eq:sigma}
&&\sigma^2\left\{\tilde P_{\Omega}(\alpha;w)\right\}\notag\\
&=&\dfrac1{N(N-1)}\sum_{j=1}^N\left[f_\Omega(w,\Lambda_{j,\alpha})-\WV{\tilde{P}_\Omega(\alpha;w)}\right]^2; 
\end{eqnarray}
}see also Ref.~\cite{Kie12}. The combined application of Eqs.~\eqref{eq:PEst} and \eqref{eq:sigma} to a given set of data immediately yields the nonclassicality quasiprobability together with its experimental errors. This yields a powerful method for the identification of quantum phenomena.

It is interesting that the pattern function in Eq.~\eqref{eq:pattern} is singular for the Gaussian filter \WV{$\Omega_w(b)=\exp\left[-b^2/2w^2\right]$} with a filter width $w\geq 1$. In the case $w=1$ the corresponding quasiprobability is the Wigner function. Consequently, even the Wigner function, which is determined in many experiments, cannot be directly sampled with a \WV{pattern function of the type under study. Instead, it can be determined by inverse Radon transform using the filtered back projection algorithm~\cite{Smi1}, by maximum likelihood methods~\cite{Hradil}, or by sampling a smoothed Wigner function~\cite{Ric1}. A strong point of our method is that it yields the nonclassicality quasiprobability locally in phase space for any point $\alpha$, which is impossible using inverse Radon transform or maximum likelihood methods. This allows us to improve the \BKTHREE{statistical} significance locally, without the need of a full state reconstruction. From this perspective, our method is much easier to implement in experiments. The resulting nonclassicality quasiprobabilities uncover general nonclassical effects, which are not directly visible in the Wigner function.}

\section{Optimal filter for experimental quantum state reconstruction}
\label{ch:analysisstandard}
There has been little research on the quality of different nonclassicality filters for experimental quantum state reconstruction.
Our purpose is to identify such nonclassicality filters 
which need a minimal number $N$ of data points in order to certify the  nonclassicality of a broad class of quantum states with the desired 
statistical significance. The significance is defined as 
\begin{equation}\label{eq:Significance}
S(N)=\left[-\inf_{\alpha,w}\dfrac{\tilde P_\Omega(\alpha;w)}{\sigma\left\{\tilde P_\Omega(\alpha;w)\right\}}\right]_{+},
\end{equation}
optimized with respect to the filter width $w$ and the phase-space point $\alpha$. The symbol $[\cdot]_{+}$ is defined as
\begin{equation}\label{eq:xposdef}
\left[x\right]_{+}=
\begin{cases}
x & \text{if }x>0,\\
0 & \text{otherwise}.
\end{cases}
\end{equation}
In the following we will apply the filters $\Omega_w^{(q)}(\beta)$ [see Eq. \eqref{eq:autq}] for various values of $q$ to several standard quantum states. The number $N$ will be determined, which is sufficient to obtain a reliable significance [Eq.~\eqref{eq:Significance}] of five standard deviations. We will refer to this number $N$ simply as the required number of data points.
\subsection{Noisy Fock state}
We begin our analysis with the simple example of a photon number state, $\hat\rho=|n\rangle\langle n|$. This state has a highly singular $P$~function containing derivatives of the $\delta$ distribution~\cite{Aga5}. In order to include effects of losses which naturally occur in experiments, the state is combined
%\WV{becomes a mixture ???} 
at a beam splitter with vacuum noise\BKTHREE{. The resulting transmitted output state has a lower mean intensity than the input state}. In the following, the notion of a noisy state is used for the resulting state of the light transmitted through the beam splitter. %This type of loss can be included in the quantum efficiency $\eta$, together with the losses due to imperfect detection.
 \BKTHREE{This type of loss can be included, together with the losses due to imperfect detection, in the total quantum efficiency $\eta.$} The noisy Fock states can be characterized by the characteristic functions of the $P$~function,
\begin{equation}
\Phi_n(\beta)=L_n(\eta|\beta|^2),
\end{equation}
where $n$ denotes the number of energy quanta and $L_n(x)$ are the Laguerre polynomials.

\begin{figure}[h]
\centerline{\includegraphics[clip,scale=0.38]{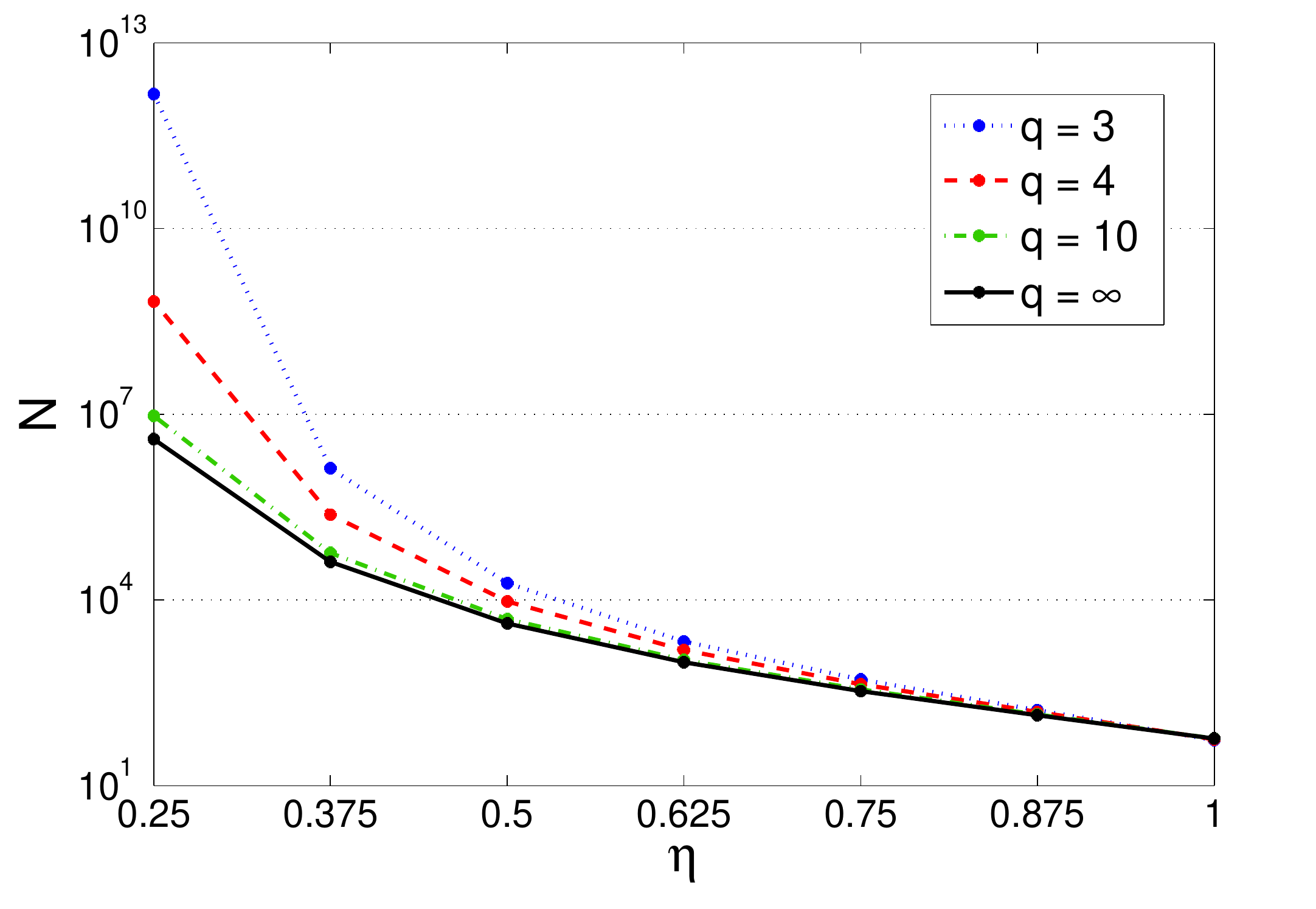}}
\vspace{0.3cm}
\centerline{\includegraphics[clip,scale=0.38]{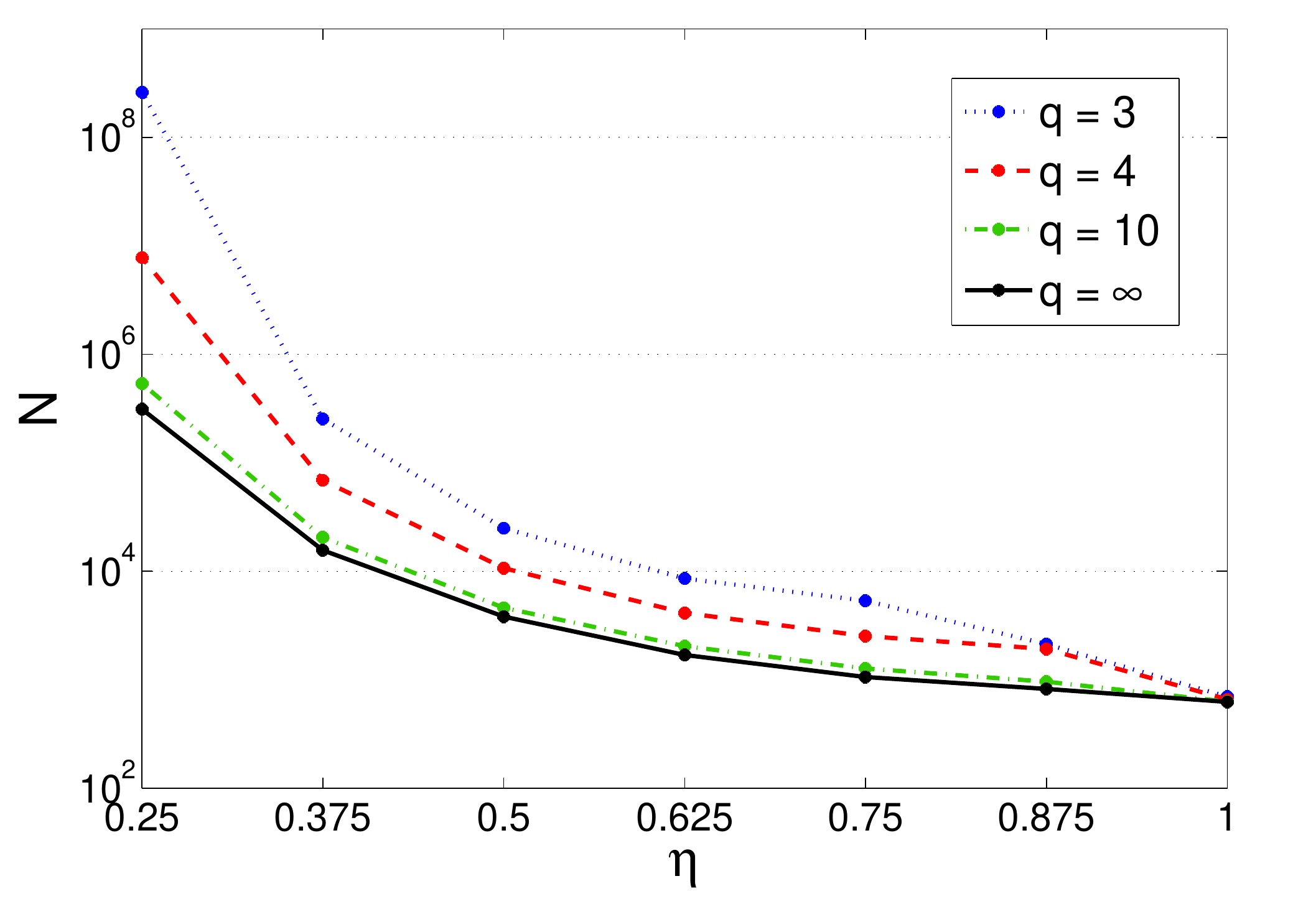}}
\caption{(Color online) Required number $N$ of data pairs $(x_j,\varphi_j)$ from BHD using the filter $\Omega_w^{(q)}(\beta)$ with different values of $q$. (top) Fock state $|1\rangle$. (bottom) Fock state $|2\rangle$. The number $N$ is calculated for various quantum efficiencies $\eta$ (dots). The filter width $w$ is optimized for each filter and each considered quantum efficiency.}
\label{fig:SPATSn0_0} 
\end{figure}
The required number $N$ of data points for sampling the regularized $P$~function from BHD data is shown in Fig.~\ref{fig:SPATSn0_0} for Fock states with $n=1$  and $2$ as a function of $\eta$ for various values of the filter parameter $q$. One observes that $N$ grows more than exponentially with decreasing quantum efficiency. For large $\eta$, $N$ depends only slightly on the filter $\Omega_w^{(q)}(\beta)$. For $\eta=0.25$, where the Wigner function is nonnegative, around $4 \times 10^5$ times more data points 
are required by using $\Omega_w^{(3)}(\beta)$ compared with $\Omega_w^{(\infty)}(\beta)$ in the case of the noisy single-photon state. For a noisy two-photon state, one needs about $8\times 10^2$ times more data points. This is an important result since, in practice, one often has limited measurement and computation time. Hence, one can save a lot of time if $\Omega_w^{(\infty)}(\beta)$ is used to certify the nonclassicality of a noisy Fock state in the case of a small quantum 
efficiency.

\subsection{Noisy multiphoton-added thermal states}
Fock states $|n\rangle$ can be considered an addition of $n$ photons to the vacuum state. Now, we want the harmonic oscillator to be initially in a thermal state described by the density operator
\begin{equation}
\hat\rho_{\mathrm{th}}=\dfrac1{\overline{n}+1}\sum_{k=0}^\infty \left(\dfrac{\overline{n}}{\overline{n}+1}\right)^k |k\rangle\langle k|.
\end{equation}
Here $\overline{n}$ is the mean thermal photon number, $\overline{n}=\mathrm{Tr}[\hat\rho_{\mathrm{th}}\hat n]$. By adding $n\geq 1$ photons to this state, we obtain a so-called $n$-photon-added thermal state ($n$-PATS):
\begin{eqnarray}
\hat\rho_{\mathrm{th}+n}&=&\mathcal{N}_n\left(\hat a^\dagger\right)^n\hat\rho_{\mathrm{th}}\hat a^n\notag\\
&=&\dfrac1{\left(\overline{n}+1\right)\,\overline{n}^n}\sum_{k=n}^\infty \binom{k}{n}\left(\dfrac{\overline{n}}{\overline{n}+1}\right)^{k} |k\rangle\langle k|, 
\end{eqnarray}
with $\mathcal{N}_n$ being the normalization constant. \WV{Single-photon-added thermal states (SPATS) have} been realized experimentally~\cite{Zav1}. Again, we include the losses in the global quantum efficiency $\eta$.

The characteristic function of the $P$~function of such a noisy $n$-photon-added thermal state reads
\begin{equation}
\Phi_n(\beta)=L_n\left(\left(1+\overline{n}\right)\eta|\beta|^2\right)e^{-\overline{n}\eta|\beta|^2}.
\label{eq:CharaNPATS}
\end{equation}
If $\overline{n}$ is larger than a certain threshold $\overline{n}_c(n)$ (Table~\ref{tab:table1}), the absolute value of the characteristic function does not exceed the value $1$. Thus, \WV{the characteristic function (CF) criterion for nonclassicality introduced} in Ref.~\cite{Vog1}, 
\begin{equation}
\label{eq:CF}
| \Phi(\beta) | > 1,
\end{equation}
which is based on Bochner's theorem of first order~\cite{Ric2,Boc1}, does not uncover nonclassicality. For related experiments, we also refer to~\cite{Zav1}.
 \begin{table}[h]
\caption{The threshold mean thermal photon number $\overline{n}_c$ depending on the number $n$ of added photons. For $\overline{n}\geq\overline{n}_c$ the absolute value of the characteristic function does not exceed the value of $1$.}
\begin{tabular}{c|cccccc}
\hline\hline
$n$ & 1 & 2 & 3 & 4 & 5 & 6\\
\hline
$\overline{n}_c$ & 0.386 & 0.549 & 0.640 & 0.698 & 0.739 & 0.770\\
\hline\hline
\end{tabular}
\label{tab:table1}
\end{table}

The Fourier transform of Eq.~\eqref{eq:CharaNPATS} yields the corresponding $P$~function~\cite{Ag-Tara},
\begin{equation}
P_n(\alpha)=\dfrac{(-1)^n}{\pi\overline{n}^{n+1}\eta}L_n\left(\dfrac{1+\overline{n}}{\overline{n}}\dfrac{|\alpha|^2}{\eta}\right) e^{-|\alpha|^2/\eta\overline{n}},
\end{equation}
which is a regular function and has negativities for all combinations of $\overline{n}$ and $\eta$, indicating that the $n$-PATS is a nonclassical state. The Wigner function of this state is given by
\begin{eqnarray}
W_n(\alpha)&=&\dfrac{2}{\pi}\dfrac{\left(1-2\eta\right)^n}{\left(1+2\overline{n}\eta\right)^{n+1}} e^{-2|\alpha|^2/\left(1+2\overline{n}\eta\right)}\notag\\
&&\times L_n\left(-\dfrac{4\left(1+\overline{n}\right)\eta|\alpha|^2}{\left(1+2\overline{n}\eta\right)\left(1-2\eta\right)}\right).
\end{eqnarray}
It attains negative values only if $\eta>1/2$. 

\WV{The Mandel $Q$~parameter is given as~\cite{Man3,Ag-Tara}}
\begin{equation}
Q_n=\dfrac{\langle:\left[\Delta \hat n\right]^2:\rangle}{\langle\hat n\rangle}=\eta\,\dfrac{\overline{n}^2(n+1)-n}{\overline{n}(n+1)+n}.
\end{equation}
Therefore, in the case
\begin{equation}
\overline{n}\geq\sqrt{\dfrac{n}{n+1}},
\end{equation}
the photon statistics of the state \WV{is of the (classical) super-Poisson type.} 
We focus here on single- and two-photon-added thermal states. For the former the mean photon number $\overline{n}$ is set to $0.8$, and for the latter it is set to $0.9$ in order to preclude the possibility of verifying nonclassicality both by the CF criterion~\eqref{eq:CF} and by a negative Mandel $Q$~parameter.

\begin{figure}[h]
\hbox{\hspace{0.4cm}\includegraphics[clip,scale=0.37]{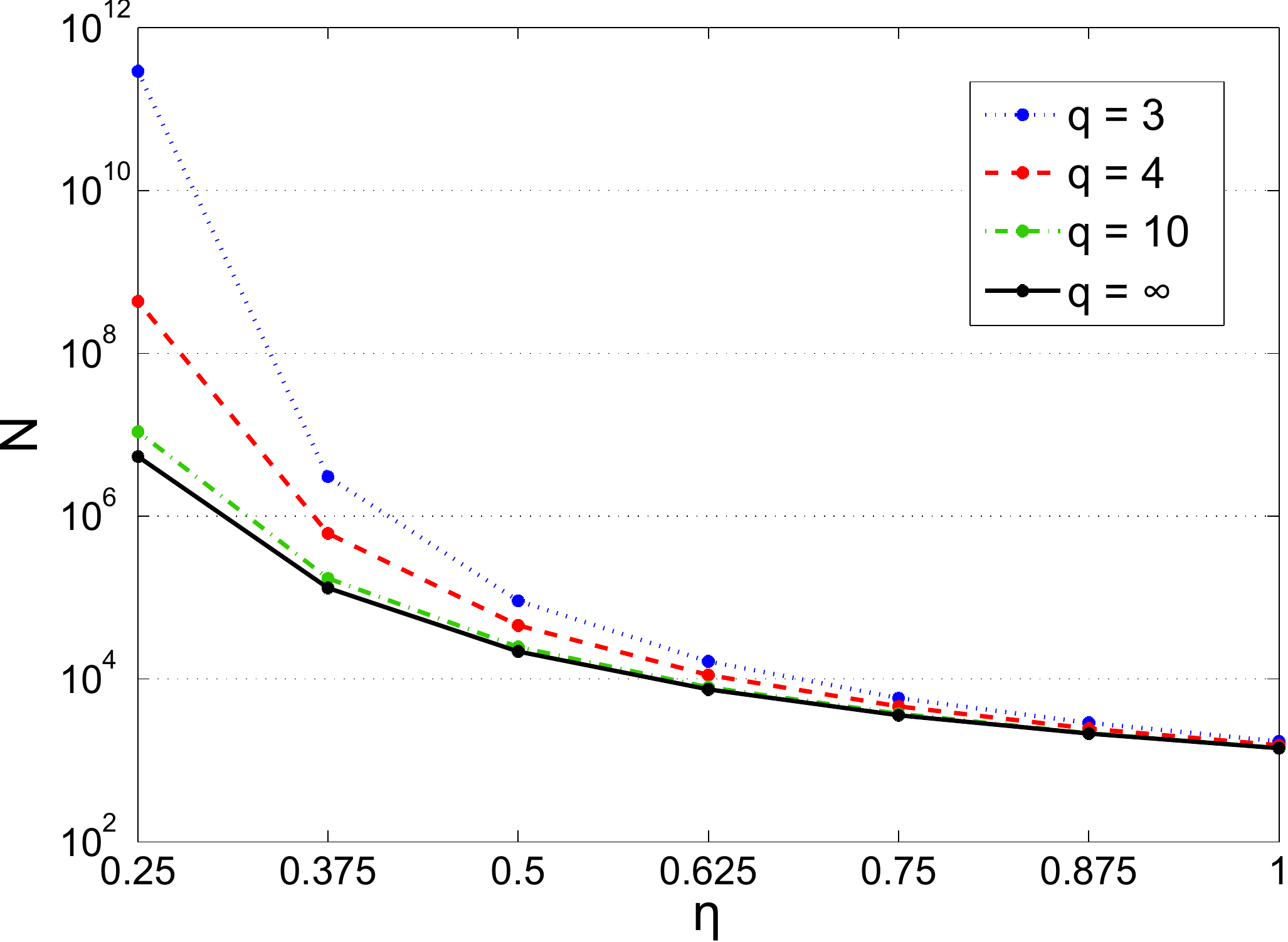}}%{autofilterlinear.png}
\vspace{0.3cm}
\hbox{\hspace{0.45cm}\includegraphics[clip,scale=0.386]{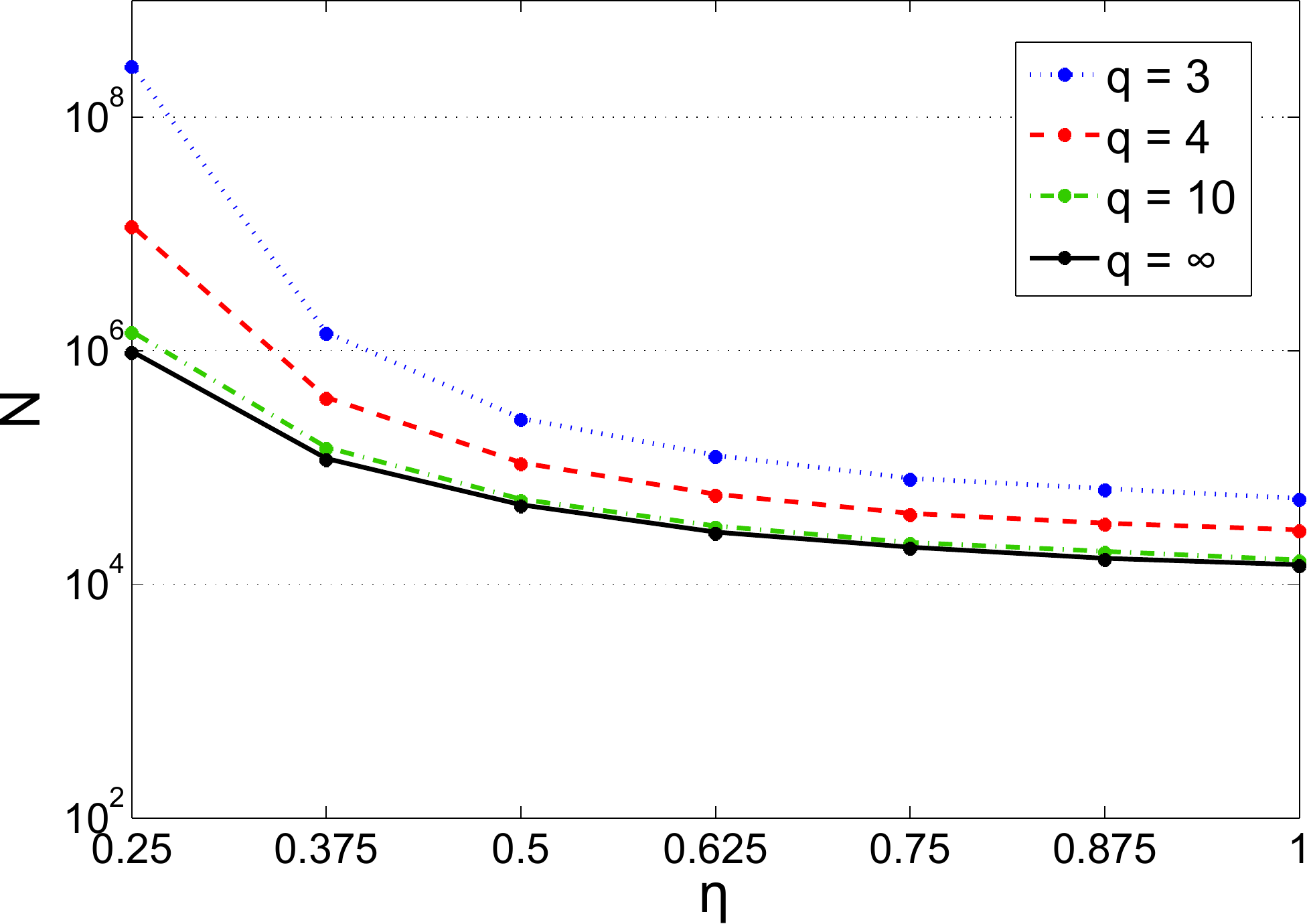}}%{autofilterlinear.png}
\caption{(Color online) Required number $N$ of data pairs $(x_j,\varphi_j)$ from BHD using the filter $\Omega_w^{(q)}(\beta)$ with different values of $q$. (top) SPATS for $\overline{n}=0.8$. (bottom) $2$-PATS for $\overline{n}=0.9$. The number $N$ is calculated for various quantum efficiencies $\eta$ (dots). The filter width $w$ is optimized for each filter and each considered quantum efficiency.}
\label{fig:PATS} 
\end{figure}
\WV{For sufficiently large numbers of data points,
the filtered $P$~functions for both states reveal statistically significant negativities, even for $\eta\leq 1/2$, where the Wigner function is nonnegative.} In Fig.~\ref{fig:PATS} the required number of data points is shown for both states. This is done for different values of the filter parameter $q$ and for various quantum efficiencies. We observe, as in the case of the noisy Fock states, that in the range of small $\eta$ much less data points are needed if the filter $\Omega_w^{(\infty)}(\beta)$ is used, compared with $\Omega_w^{(3)}(\beta)$ and $\Omega_w^{(4)}(\beta)$.

\subsection{Completely dephased squeezed vacuum state}
As a third example for a quantum state whose nonclassicality is difficult to verify by standard methods, we consider a squeezed vacuum state, whose phase is completely randomized. As for the previous states, we include losses in the quantum efficiency $\eta$. This state, which is a mixture of squeezed states, has a nonnegative Wigner function and does not exhibit squeezing. Moreover, it has a highly singular $P$~function which cannot be determined directly from experiments. The origin of the singularity of $P(\alpha)$ is seen from its characteristic function, 
\begin{equation}\label{eq:SVchara}
\Phi(\beta)=\exp\left[-\dfrac{\eta|\beta|^2}{4}(V_x+V_p-2)\right]\,\mathrm{I}_0\left(\dfrac{\eta|\beta|^2}{4}(V_x-V_p)\right),
\end{equation}
which tends to infinity for $|\beta|\to\infty$. Here $V_x$ and $V_p$ are the orthogonal quadrature variances and $\mathrm{I}_0(x)$ is the modified Bessel function of the first kind.

\begin{figure}[h]
\includegraphics[clip,scale=0.38]{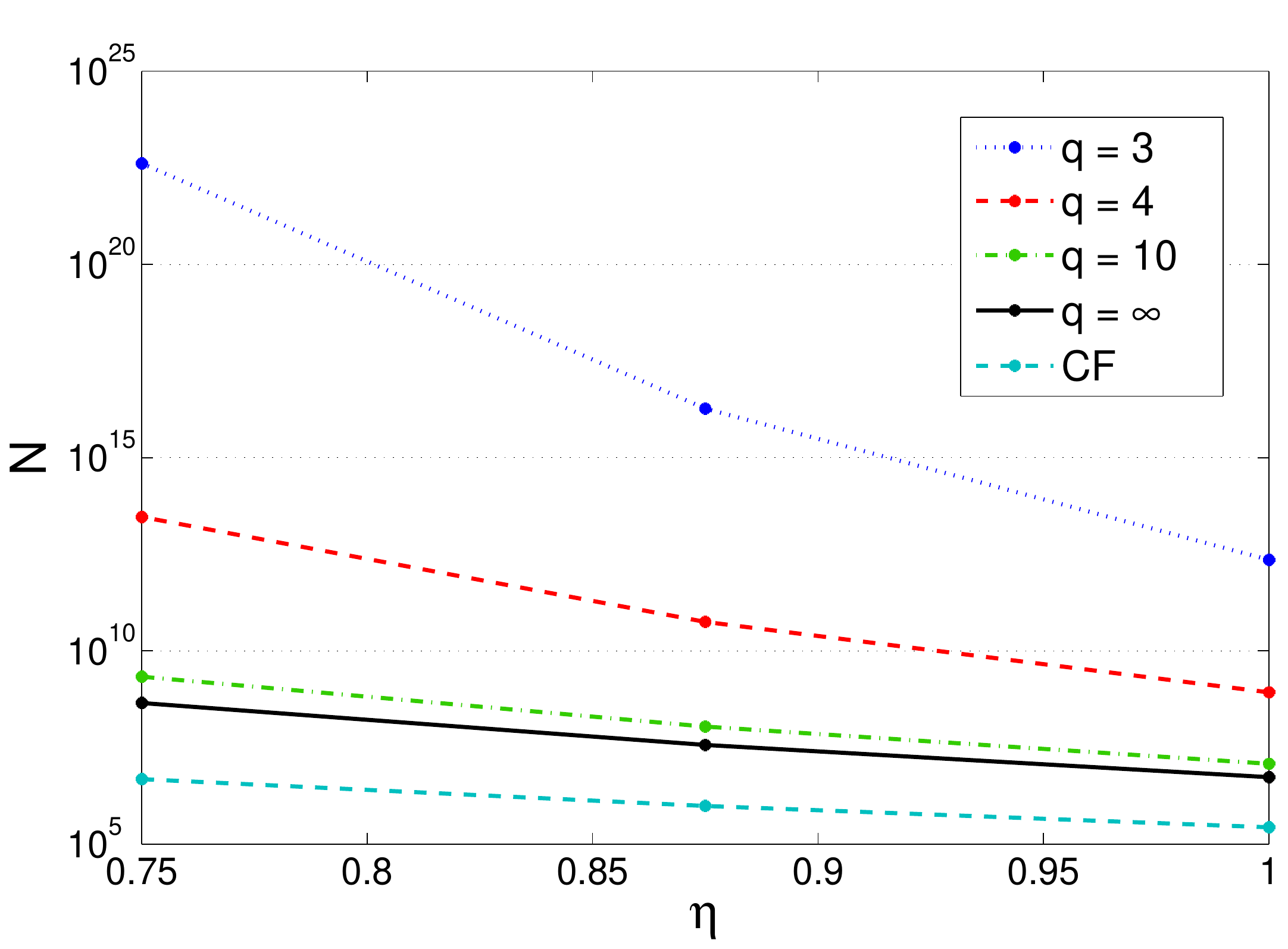}%{autofilterlinear.png}
\caption{(Color online) Required number $N$ of data pairs $(x_j,\varphi_j)$ from BHD using the filter $\Omega_w^{(q)}(\beta)$ with different values of $q$. The state under study is a completely phase randomized squeezed vacuum state with orthogonal quadrature variances $V_x=0.4$ and $V_p=5.0$. The number $N$ is calculated for various quantum efficiencies $\eta$ (dots). The filter width $w$ is optimized for each filter and each considered quantum efficiency. The line \WV{labeled CF shows the number of data points for a statistical significance of five standard deviations} for the negativity of $1-|\Phi(\beta)|$.}
\label{fig:SV} 
\end{figure}
The characteristic function~\eqref{eq:SVchara} exceeds the value of $1$ only for relatively large values of $|\beta|$. However, the number of data points needed to certify 
nonclassicality significantly by using the CF criterion~\eqref{eq:CF} (see~\cite{Vog1,Lvo1}) grows exponentially with increasing $|\beta|$~\cite{Kie80}. Hence, we compare the required number of data points for this criterion with those for the different nonclassicality filters $\Omega_w^{(q)}(\beta)$.
The result is shown in Fig.~\ref{fig:SV}. The filter $\Omega^{(\infty)}_w(\beta)$ needs many orders of magnitude fewer data points than $\Omega^{(3)}_w(\beta)$ and $\Omega^{(4)}_w(\beta)$, even for $\eta=1$. However, a significant verification of nonclassicality via the CF criterion is still possible with about two orders of magnitude fewer data points, compared with a nonclassicality test based on the filter $\Omega^{(\infty)}_w(\beta)$.

\begin{figure*}[ht]
\includegraphics[scale=0.46]{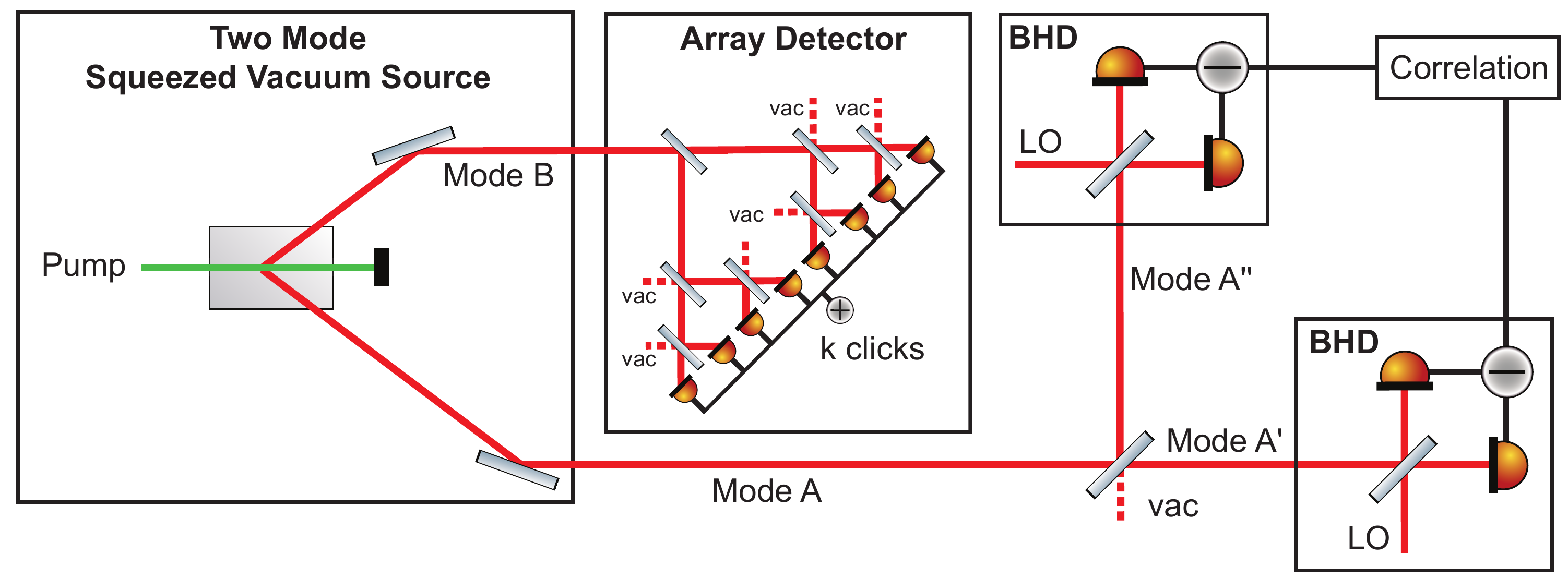}
\caption{(Color online) \BK{The light source prepares a two-mode squeezed vacuum state. The quantum state of mode $A$ is prepared, conditioned on the events recorded in mode $B$ by an array of $M=8$ on-off detectors. Mode $A$ is subdivided by a beam splitter into modes $A'$ and $A''$, whose correlations are measured by two BHDs.}}
%Experimental setup for the heralded state preparation. The detector array consists of $M=8$ on-off detectors.
\label{fig:SetupHerald}
\end{figure*}
\begin{figure*}[ht]
 \includegraphics[width=0.41\textwidth]{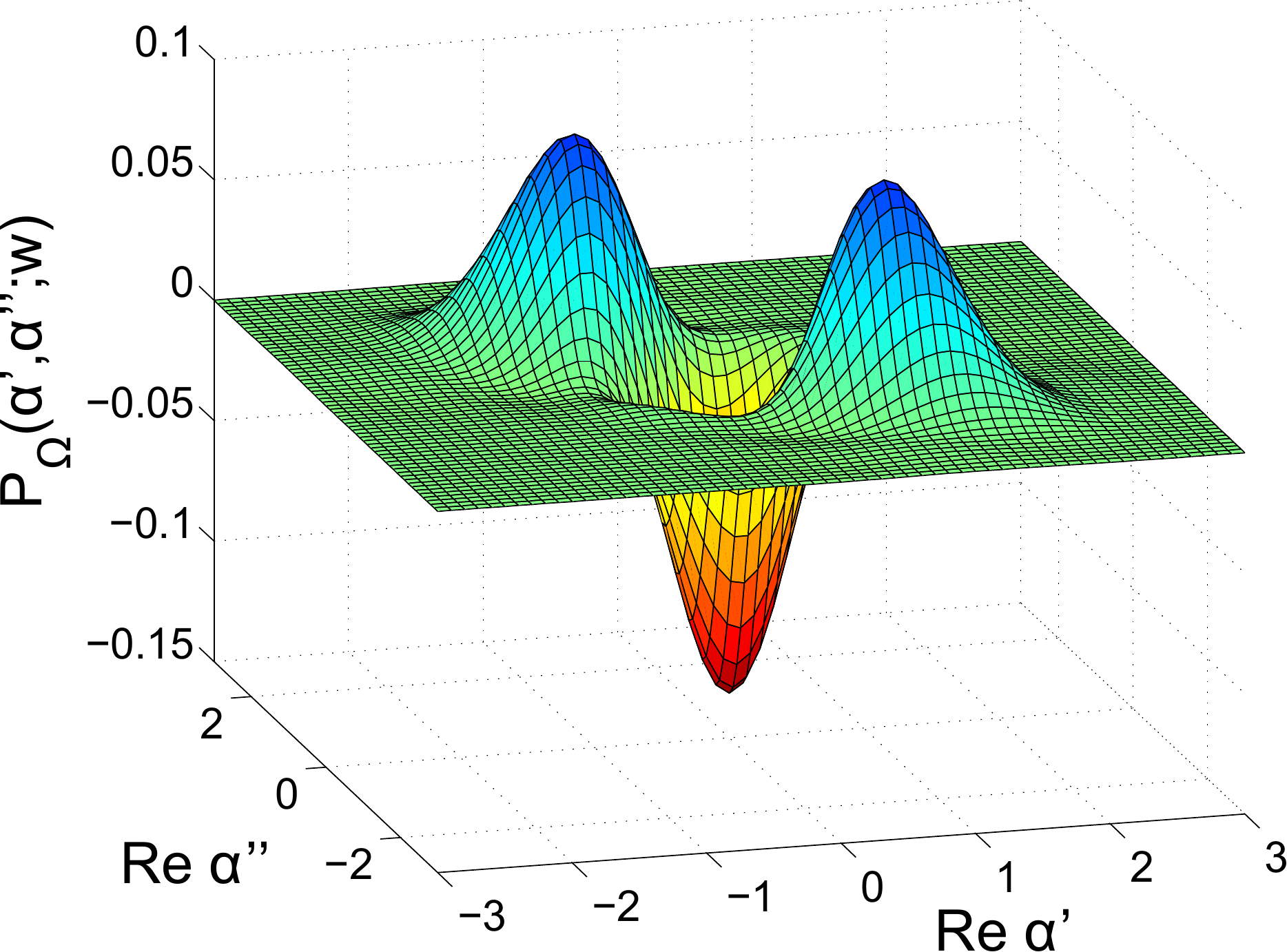}
 %\vspace{0.1cm}
\hspace{0.5cm}
 \includegraphics[width=0.41\textwidth]{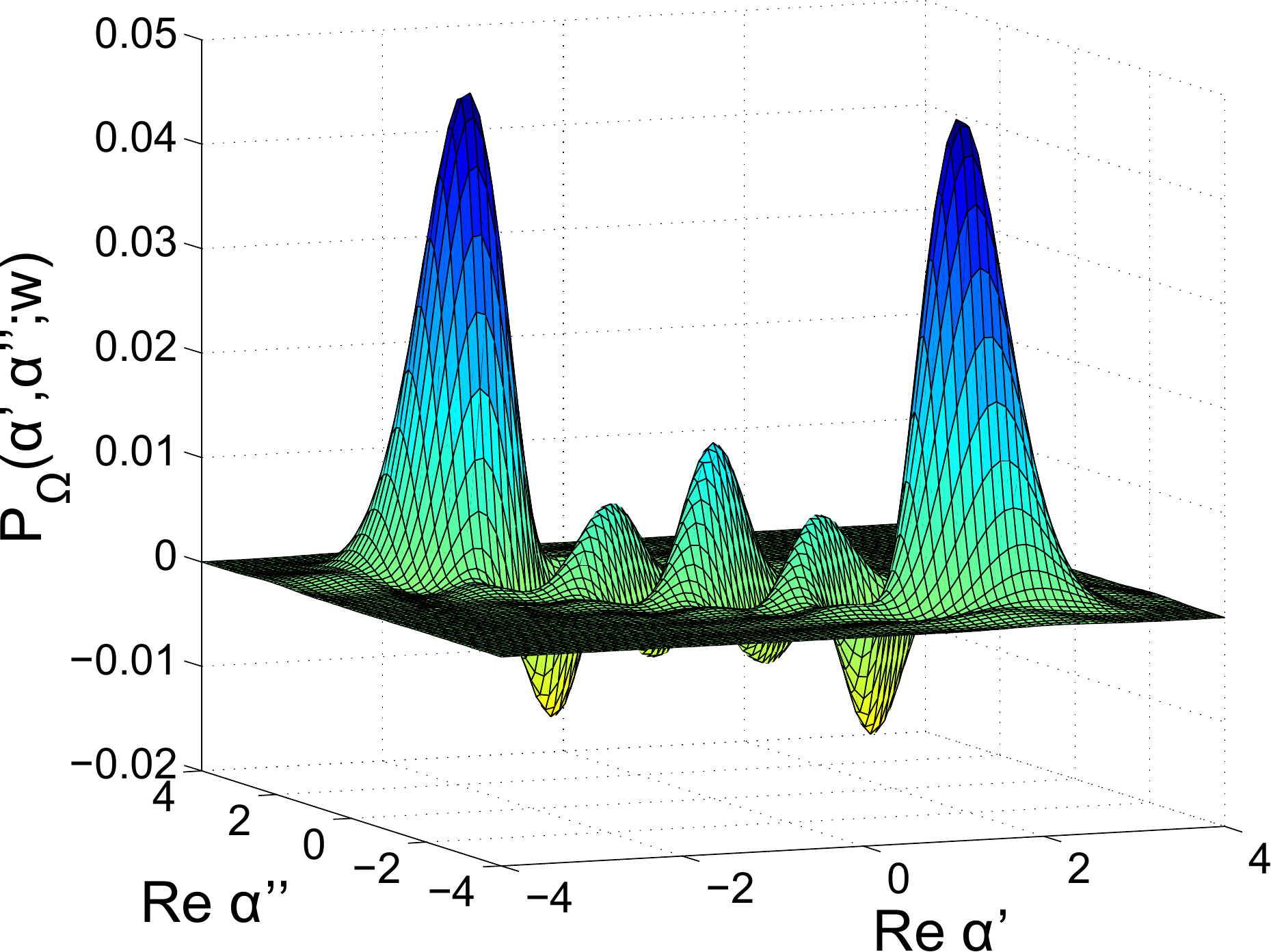}
 \caption{(Color online) Regularized $P$~function \BK{$P_\Omega(\alpha ',\alpha '';w)$} of the two\BKTWO{-}mode system, modes $A'$ and $A''$, for (left) k=1  and (right) k=4 clicks being recorded by the array detector. The filter $\Omega_w^{(\infty)}(\beta)$ with a filter width $w=1.7$ is applied. Distinct negativities in both cases visualize the nonclassical correlations of the states.}
\label{fig:tmk}
\end{figure*}

\section{Multimode nonclassicality quasiprobability}
\label{ch:twomode} 
Here we will show that %nonclassicality 
regularized $P$~functions can also visualize the nonclassicality of multipartite systems. %can also be visualized by regularized $P$~functions \WV{of multimode systems.
Let us consider a heralded state engineering scenario as illustrated in Fig.~\ref{fig:SetupHerald}. A source prepares the light modes $A$ and $B$ in a two\BKTWO{-}mode squeezed vacuum state~\cite{Ebe1}, which reads, in the photon number basis,
\begin{equation}
\hat\rho_{\mathrm{in}}=\sum_{p,q=0}^\infty\mathrm e^{i(p-q)\arg(\xi)}\dfrac{\left(-\tanh|\xi|\right)^{p+q}}{\cosh^2|\xi|}|p,p\rangle\langle q,q|.
 \end{equation}
The parameter $\xi$ describes the amount of squeezing. Mode $B$ undergoes a measurement with a so-called array detector~\cite{Spe2}. This detector consists of a cascaded arrangement of 50:50 beam splitters dividing the light beam $B$ into $M$ output beams, each measured with an on-off detector with quantum efficiency $\eta$~\cite{Jia1}. Conditioned on the detection of $k$ clicks by the array detector, the resulting state of mode $A$ is given by the $P$~function
 \begin{eqnarray}\label{eq:HeraldPFunction} P(\alpha)&=&\mathcal{N}_P\sum_{j=0}^k\binom{k}{j}\dfrac{(-1)^j}{\zeta(1-\eta+\eta j/M)}\notag\\
&&\times\exp\left[-\left(\dfrac1{\zeta(1-\eta+\eta j/M)}-1\right)|\alpha|^2\right],\notag\\ 
\end{eqnarray}
as can be derived by applying the methods given in Ref.~\cite{Spe1}. The constant $\mathcal{N}_P$ ensures correct normalization and $\zeta=\tanh^2|\xi|$.

Mode $A$ is subdivided by a beam splitter into two beams referred to as modes $A'$ and $A''$. %if both output modes $A'$ and $A''$ in Fig.~\ref{fig:SetupHerald} are considered. 
\WV{The fraction of the intensity of mode $A$ transmitted into mode $A'$ is 
$\eta_L$.} The density operator of the two\BKTWO{-}mode system, modes $A'$ and $A''$, in coherent-state basis reads
\begin{equation}
\hat\rho=\int d^2\alpha '\,d^2\alpha '' P(\alpha ',\alpha '')|\alpha '\rangle\langle \alpha '|\otimes|\alpha ''\rangle\langle\alpha ''|,
\end{equation} 
where the two\BKTWO{-}mode $P$~function is given by
\begin{equation}\label{eq:twomodeP}
P(\alpha ',\alpha '')=\dfrac1{\eta_L}\,P\left(\dfrac{\alpha '}{\sqrt{\eta_L}}\right)\,\delta\left(\dfrac{\sqrt{1-\eta_L}}{\sqrt{\eta_L}}\alpha '+\alpha ''\right)
\end{equation}
(see, e.g.,~\cite{Kie10}). %The function $P(\alpha)$ is the $P$~function of the beam splitter input, mode $A$, as derived in Eq.~\eqref{eq:HeraldPFunction}.
The function $P(\alpha)$ is the $P$~function of the beam-splitter input mode $A$, given in Eq.~\eqref{eq:HeraldPFunction}.
The other input is a vacuum state whose $P$~function is a $\delta$~function in the origin of the phase space. We consider a 50:50 beam splitter, which implies that $\eta_L=1/2$. 

The two-mode $P$~function, Eq.~\eqref{eq:twomodeP}, cannot be obtained from experiments since it contains a singular $\delta$~function. It is, however, possible to generalize the method of nonclassicality filters to multimode systems~\cite{Agu1}. For the considered two\BKTWO{-}mode system the nonclassicality quasiprobability \BK{$P_\Omega(\alpha ',\alpha '';w)$} is obtained from the $P$~function via
\begin{eqnarray}
\BK{P_\Omega(\alpha ',\alpha '';w)}&=&\int d^2\beta '\,d^2\beta ''\,P(\beta ',\beta '')\,\mathcal{F}\left[\Omega_w\right](\alpha '-\beta ')\notag\\
&&\times\,\mathcal{F}\left[\Omega_w\right](\alpha ''-\beta '').
\end{eqnarray}
Here, $\mathcal{F}\left[\Omega_w\right](\alpha)$ is the Fourier transform of the single\BKTWO{-}mode nonclassicality filter, with filter width $w$. 

Figure~\ref{fig:tmk} shows a subspace of the regularized two\BKTWO{-}mode $P$~functions 
for two different numbers $k$ of clicks of the array detector, and the filter  $\Omega_w^{(\infty)}(\beta)$. The squeezing parameter $\xi$ is chosen to be $1.0$ and the on-off detector efficiency $\eta$ is set to $\eta=0.2$. The
 regularized two\BKTWO{-}mode functions clearly attain negative values, which certify nonclassical correlations of both two-mode states. \WV{For an increasing number of clicks, more quantum interference structures occur in the two-mode state.}

\section{Summary and Conclusions}
\label{ch:conclusion}
\WV{An important class of nonclassicality filters $\Omega_w^{(q)}(\beta)$, parametrized by two real parameters $q$ and $w$, has been analyzed with respect to its practical applicability in experiments. These filters are based on the autocorrelation functions of rapidly decaying functions, with the decay being controlled by the value of $q$ and the filter width $w$. 
In principle, they can be applied to any quantum state to uncover all types of nonclassical effects.}
%\begin{figure*}[ht]

\WV{We have shown that an optimal experimental filtering with these filters is possible in the limit of $q\to \infty$. For this particular case we derive an analytical expression for the filter function, which simplifies its practical application. In this case, the minimal number of experimental data points needed to significantly vi\BKTWO{s}ualize nonclassical effects reduces substantially. More generally, it turns out that the required number of data points decreases as the $q$~value increases. Even though the $q\to \infty$ filter leads to some loss of information on the state under study for any finite value of the filter width $w$,
it still \BKTWO{asymptotically} approaches the full information when the value of $w$ is increased. As the choice of this value in practice is intimately related to the noise contained in the experimental data, this loss of information is merely caused by the statistical errors in the experiment, rather than being a fundamental limitation.}

\WV{The advantages of the filter with $q\to \infty$ increase dramatically for decreasing quantum efficiencies. In this range, the number of data points needed to
verify quantum effects with a desired significance reduces by orders of magnitude.
This observation was made for different types of quantum states, such as noisy Fock states, $n$-photon\BKTWO{-}added thermal states, and fully dephased squeezed \BKTWO{vacuum} states.
We have also studied the application of this nonclassicality filter to multimode radiation fields. As an example, we have derived the nonclassicality quasiprobabilities 
for two-mode fields, prepared in quantum-correlated states by heralding methods with an array detector.}

\WV{Based on a theorem on positive-definite functions by Askey, we could derive an alternative type of nonclassicality filter in an analytical form. 
The properties of this filter are not optimal for the application in experiments, since the statistical noise of the data is not suppressed sufficiently well. 
However, this filter is very useful for analyzing quantum effects 
in theory or for theoretical simulations of experiments.}

\begin{acknowledgements}
This work was supported by the Deutsche Forschungsgemeinschaft through SFB 652. The authors gratefully acknowledge stimulating discussions with J. Sperling.
\end{acknowledgements}
\section*{Appendix: Proof of Conditions 1 and 2 for the analytical invertible filter}
\appendix*
\setcounter{equation}{0}
We prove \textit{Condition 1} for the filter $\Omega_w(\beta;s,C)$ by calculating the squared $L^2$ norm,
\begin{eqnarray}
&&\left(\left\|\Omega_w(\beta;s,C)e^{|\beta|^2/2}\right\|_2\right)^2\notag\\
&=&\left(\left\|\exp\left[-\left(\dfrac{|\beta|}{w}+C\right)^s+C^s\right]e^{|\beta|^2/2}\right\|_2\right)^2\notag\\
&=&\int d^2\beta \exp\left[-2\left(\dfrac{|\beta|}{w}+C\right)^s+2C^s\right]e^{|\beta|^2}\notag\\
&=&2\pi\exp\left(2C^s\right)\int_0^\infty db\,b\,\exp\left[-2\left(\dfrac{b}{w}+C\right)^s\right]e^{b^2}.\notag\\
\end{eqnarray}
Using that 
\begin{equation}
\exp\left[-2\left(\dfrac{b}{w}+C\right)^s\right]<\exp\left[-2\left(\dfrac{b}{w}\right)^s\right] 
\end{equation}
(note that $C>0$), we find
\begin{eqnarray}
&&\left(\left\|\Omega_w(\beta;s,C)e^{|\beta|^2/2}\right\|_2\right)^2\notag\\
&<&2\pi\exp\left(2C^s\right)\int_0^\infty db\,b\,\exp\left[-2\left(\dfrac{b}{w}\right)^s+b^2\right].\notag\\
\end{eqnarray}
Now, we split this integral into two integrals. By defining $a=\sqrt[s-2]{w^s}$, we derive 
\begin{eqnarray}\label{eq:squareintana}
&&\left(\left\|\Omega_w(\beta;s,C)e^{|\beta|^2/2}\right\|_2\right)^2\notag\\
&<&2\pi\exp\left(2C^s\right)\int_0^{a} db\,b\,\exp\left[-2\left(\dfrac{b}{w}\right)^s+b^2\right]\notag\\
&&+2\pi\exp\left(2C^s\right)\int_{a}^\infty db\,b\,\exp\left[-2\left(\dfrac{b}{w}\right)^s+b^2\right]\notag\\
&=&K+2\pi\exp\left(2C^s\right)\int_{a}^\infty db\,b\,\exp\left[-b^2\left(2\dfrac{b^{s-2}}{w^s}-1\right)\right],\notag\\
\end{eqnarray}
with $K<\infty$. The integration variable $b$ is, due to the integration limits, greater than $\sqrt[s-2]{w^s}$. If $s>2$, then 
\begin{equation}
2\dfrac{b^{s-2}}{w^s}-1\geq 1
\end{equation}
holds and accordingly, the inequality
\begin{equation}
\exp\left[-b^2\left(2\dfrac{b^{s-2}}{w^s}-1\right)\right]\leq\exp\left[-b^2\right]
\end{equation}
is fulfilled. This allows us to further estimate~\eqref{eq:squareintana}:
\begin{eqnarray}
&&\left(\left\|\Omega_w(\beta;s,C)e^{|\beta|^2/2}\right\|_2\right)^2\notag\\
&<&K+2\pi\exp\left(2C^s\right)\int_{a}^\infty db\,b\,\exp\left[-b^2\right]\notag\\
&\leq&K+2\pi\exp\left(2C^s\right)\int_0^\infty db\,b\,\exp\left[-b^2\right]\notag\\
%&=&K+\pi\mathrm e^{2C^s}\int_0^\infty dz\,\mathrm{e}^{-z}\notag\\
&=&K+\pi\exp\left(2C^s\right)\notag\\
&<&\infty.
\end{eqnarray}
~$\hfill\Box$\\\\
For the verification of \textit{Condition 2}, Askey's theorem, which is given in Sec.~\ref{sec:anafilter}, is used. The function $\Omega_w(\beta;s,C)$ depends on two real variables $\rm{Re}(\beta)$ and $\rm{Im}(\beta)$, due to the dimension of the phase space of a one-dimensional harmonic oscillator. Since we demand that its Fourier transform is nonnegative, we have to ensure that the conditions of the theorem are satisfied for $n=2$. Hence, we have to prove that
\begin{equation}
-\dfrac{d}{d|\beta|}\Omega_w(\beta;s,C)
\end{equation}
is convex for all $|\beta|$, i.e., 
\begin{equation}\label{eq:thirdderivative4}
\forall\,|\beta|:\,\,-\dfrac{d^3}{d|\beta|^3}\Omega_w(\beta;s,C)\geq 0.
\end{equation}
Straightforward calculation yields
\begin{eqnarray}
&&-\dfrac{d^3}{d|\beta|^3}\Omega_w(\beta;s,C)\notag\\
&=&-\left[-s(s-1)(s-2)+3s^2(s-1)\left(\dfrac{|\beta|}{w}+C\right)^s\right.\notag\\
&&\left.-s^3\left(\dfrac{|\beta|}{w}+C\right)^{2s}\right]\underbrace{\dfrac1{w^3}\left(\dfrac{|\beta|}{w}+C\right)^{s-3}}_{>0}\underbrace{\Omega_w(\beta;s,C)}_{>0}.\notag\\
\end{eqnarray}
Therefore,~\eqref{eq:thirdderivative4} is fulfilled if
\begin{eqnarray}
&&\left[-s(s-1)(s-2)+3s^2(s-1)\left(\dfrac{|\beta|}{w}+C\right)^s\right.\notag\\
&&\left.-s^3\left(\dfrac{|\beta|}{w}+C\right)^{2s}\right]\leq 0
\end{eqnarray}
for all $|\beta|$. This is true if
\begin{equation}\label{eq:Cwb}
\forall\,|\beta|:\,\,\dfrac{|\beta|}{w}\geq -C+\left(\dfrac{3(s-1)+\sqrt{1-6s+5s^2}}{2s}\right)^{1/s}.
\end{equation}
Consequently, the right hand side of~\eqref{eq:Cwb} has to be smaller than or equal to zero, requiring that the parameter $C$ fulfills
\begin{equation}\label{eq:Cs}
C\geq\left(\dfrac{3(s-1)+\sqrt{1-6s+5s^2}}{2s}\right)^{1/s}.
\end{equation}
~$\hfill\Box$\\\\


\begin{thebibliography}{999pt}
\bibitem{Nielsen}M. A. Nielsen and I. L. Chuang, {\it Quantum Computation and Quantum Information}, (Cambridge University Press, Cambridge, 2000).
\bibitem{HorodeckiRev}R. Horodecki, P. Horodecki, M. Horodecki, and K. Horodecki, Rev. Mod. Phys. \textbf{81}, 865 (2009).
%\bibitem[1]{Lia1}L. Jiang, J. M. Taylor, A. S. S\o rensen, M. D. Lukin \textcolor{blue}{Phys. Rev. A \textbf{76}, 062323 (2007)}.
\bibitem{Ben1}C. H. Bennett, G. Brassard, C. Cr\'{e}peau, R. Jozsa, A. Peres, and W. K. Wootters, Phys. Rev. Lett. \textbf{70}, 1895 (1993).
\bibitem{ahar}Y. Aharonov, D. Falkoff, E. Lerner, and H. Pendleton, Ann. Phys. (N.Y.) \textbf{39}, 498 (1966).
\bibitem{kim}M. S. Kim, W. Son, V. Bu\v{z}ek, and P. L. Knight, Phys. Rev. A \textbf{65}, 032323 (2002).
\bibitem{xiangbin}Wang Xiang-bin, Phys. Rev. A \textbf{66}, 024303 (2002).
\bibitem{WEP03}M. M. Wolf, J. Eisert, and M. B. Plenio, Phys. Rev. Lett. \textbf{90}, 047904 (2003).
\bibitem{JLC13}Z. Jiang, M. D. Lang, and C. M. Caves, Phys. Rev. A \textbf{88}, 044301 (2013).
%\bibitem[3]{Kim1}M. S. Kim, W. Son, V. Bu\v{z}ek, \textcolor{blue}{Phys. Rev. A \textbf{65}, 032323 (2002)}.
\bibitem{Gla1}R. J. Glauber, Phys. Rev. \textbf{131}, 2766 (1963).
\bibitem{Sud1}E. C. G. Sudarshan, Phys. Rev. Lett. \textbf{10}, 277 (1963).
\bibitem{Tit1}U. M. Titulaer and R. J. Glauber, Phys. Rev. \textbf{140}, B676 (1965).
\bibitem{Man1}L. Mandel, Phys. Scr. \textbf{T12}, 34 (1986).
\bibitem{Geh1}C. Gehrke, J. Sperling, and W. Vogel, Phys. Rev. A \textbf{86}, 052118 (2012).
\bibitem{Ag-Tara}G. S. Agarwal and K. Tara, Phys. Rev. A \textbf{46}, 485 (1992).
\bibitem{Kie2}T. Kiesel, W. Vogel, V. Parigi, A. Zavatta, and M. Bellini, Phys. Rev. A \textbf{78}, 021804 (2008). 
\bibitem{Kie1}T. Kiesel and W. Vogel, Phys. Rev. A \textbf{82}, 032107 (2010).
\bibitem{Ric2}Th. Richter and W. Vogel, Phys. Rev. Lett. \textbf{89}, 283601 (2002).
\bibitem{Boc1}S. Bochner, Math. Ann. \textbf{108}, 378 (1933).
\bibitem{Ag} G. S. Agarwal, Opt. Commun. \textbf{95}, 109 (1993).
\bibitem{Shc1}E. V. Shchukin and W. Vogel, Phys. Rev. A \textbf{72}, 043808 (2005). 
\bibitem{Dal1}R. Short and L. Mandel, Phys. Rev. Lett. \textbf{51}, 384 (1983).   %B. J. Dalton, \textcolor{blue}{Phys. Scr. \textbf{T12}, 43 (1986)}.
\bibitem{Slusher}R. E. Slusher, L. W. Hollberg, B. Yurke, J. C. Mertz, and J. F. Valley, Phys. Rev. Lett. \textbf{55}, 2409 (1985).
\bibitem{Lin1} L.-A. Wu, H. J. Kimble, J. L. Hall, and H. Wu, Phys. Rev. Lett. \textbf{57}, 2520 (1986).
\bibitem{Wineland}D. Leibfried, D. M. Meekhof, B. E. King, C. Monroe, W. M. Itano, and D. J. Wineland,
Phys. Rev. Lett. \textbf{77}, 4281 (1996).
%K. E. Cahill and R. J. Glauber, \textcolor{blue}{Phys. Rev. \textbf{177}, 1857 (1969)}.
\bibitem{Kie5}T. Kiesel, W. Vogel, M. Bellini, and A. Zavatta, Phys. Rev. A \textbf{83}, 032116 (2011).
\bibitem{Kie-Schn}T. Kiesel, W. Vogel, B. Hage, and R. Schnabel,
%      {\em Direct Sampling of Negative Quasiprobabilities of a Squeezed State}\\
      Phys. Rev. Lett. \textbf{107}, 113604 (2011).
\bibitem{Kie-Polz} T. Kiesel, W. Vogel, S. L. Christensen, J.-B. B\'{e}guin, J. Appel, and E. S. Polzik, 
%      {\em Atomic nonclassicality quasiprobabilities}\\
      Phys. Rev. A \textbf{86}, 042108 (2012).
\bibitem{Cah1}K. E. Cahill and R. J. Glauber, Phys. Rev. \textbf{177}, 1857 (1969).      
      %\bibitem{Rah1}S. Rahimi-Keshari, T. Kiesel, and W. Vogel, \textcolor{blue}{Phys. Rev. A \textbf{85}, 043827 (2012)}.
\bibitem{Kie4}T. Kiesel and W. Vogel, Phys. Rev. A \textbf{85}, 062106 (2012).
\bibitem{Ask1}R. Askey, Mathematics Research Center, University of Wisconsin--Madison, Technical Report No. 1262, 1973 (unpublished).
\bibitem{Yue1}J. H. Shapiro, H. P. Yuen, and J. A. Machado Mata, IEEE Trans. Inf. Theory \textbf{25}, 179 (1979).
\bibitem{We99}D.-G. Welsch, W. Vogel, and T. Opatrn\'{y},
%{\it Homodyne detection and quantum-state reconstruction}
Prog. Opt. \textbf{39}, 63 (1999).
\bibitem{Kie12} \WV{T. Kiesel, Phys. Rev. A \textbf{85}, 052114 (2012)}.
% \bibitem{Man2}L. Mandel, \textcolor{blue}{Phys. Rev. Lett. \textbf{49}, 2 (1982)}.
% \bibitem{Car1}H. J. Carmichael, \textcolor{blue}{J. Opt. Soc. Am. B\textbf{4}, 1588 (1987)}.
% \bibitem{Vog2}W. Vogel and J. Grabow, \textcolor{blue}{Phys. Rev. A \textbf{47}, 4227 (1993)}.
%\bibitem{Kie3}T. Kiesel, W. Vogel, B. Hage, and R. Schnabel, \textcolor{blue}{Phys. Rev. A \textbf{107}, 113604 (2011)}.
\bibitem{Smi1}D. T. Smithey, M. Beck, M. G. Raymer and A. Faridani, Phys. Rev. Lett. \textbf{70}, 1244 (1993).
\bibitem{Hradil}Z. Hradil, Phys. Rev. A \textbf{55}, R1561 (1997). 
\bibitem{Ric1}Th. Richter, J. Opt. B \textbf{1}, 650-654 (1999).
\bibitem{Aga5}G. S. Agarwal and E. Wolf, Phys. Rev. D \textbf{2}, 2161 (1970).
\bibitem{Zav1}A. Zavatta, V. Parigi, and M. Bellini, Phys. Rev. A \textbf{75}, 052106 (2007). 
\bibitem{Vog1}W. Vogel, Phys. Rev. Lett. \textbf{84}, 1849 (2000). 
%\bibitem{Aga1}G. S. Agarwal and K. Tara, \textcolor{blue}{Phys. Rev. A \textbf{46}, 485 (1992)}.
\bibitem{Man3}L. Mandel, Opt. Lett. \textbf{4}, 205 (1979).
\bibitem{Lvo1}A. I. Lvovsky and J. H. Shapiro, Phys. Rev. A \textbf{65}, 033830 (2002).
\bibitem{Kie80}T. Kiesel, W. Vogel, B. Hage, J. DiGuglielmo, A. Samblowski, and R. Schnabel, Phys. Rev. A \textbf{79}, 022122 (2009).
%\bibitem{Kie20}T. Kiesel and W. Vogel, \textcolor{blue}{Phys. Rev. A \textbf{86}, 032119 (2012)}.
\bibitem{Ebe1}T. Eberle, V. H\"andchen, and R. Schnabel, Opt. Express \textbf{21}, 11546 (2013).
\bibitem{Spe2}J. Sperling, W. Vogel, and G. S. Agarwal, Phys. Rev. A \textbf{85}, 023820 (2012).
\bibitem{Jia1}L. A. Jiang, E. A. Dauler, and J. T. Chang, Phys. Rev. A \textbf{75}, 062325 (2007).
\bibitem{Spe1}J. Sperling, W. Vogel, and G. S. Agarwal, Phys. Rev. A \textbf{89}, 043829 (2014). 
\bibitem{Kie10}T. Kiesel, Phys. Rev. A \textbf{87}, 062114 (2013).
\bibitem{Agu1}E. Agudelo, J. Sperling, and W. Vogel, Phys. Rev. A \textbf{87}, 033811 (2013). 

\end{thebibliography}
\end{document}